\documentstyle[amssymb,twocolumn,epsf,aps]{revtex}

\begin{document}
\title{Orbital effect of in-plane magnetic field on quantum transport in chaotic
lateral dots}
\author{Vladimir I. Fal'ko$^{1}$ and Tomas Jungwirth$^{2,3}$}
\address{$^{1}$ Physics Department, Lancaster University, LA1 4YB, Lancaster, UK\\
$^{2}$ Institute of Physics, Czeck Academy of Sciences, Prague, Czeck
Republik \\
$^{3}$ Physics Department, University of Texas, Austin, USA}
\date{today}
\maketitle

\begin{abstract}
We show how the in-plane magnetic field, which breaks time-reversal and
rotational symmetries of the orbital motion of electrons in a
heterostructure due to the momentum-dependent inter-subband mixing, affects
weak localisation correction to conductance of a large-area chaotic lateral
quantum dot and parameteric dependences of universal conductance
fluctuations in it.
\end{abstract}


A high sensitivity of phase-coherent transport through quantum dots to
external perturbations has recently enabled one to transform studies of
mesoscopic effects \cite{Washburn,ChemicalPot,AltshulerKhm} into a
spectroscopic tool for detecting tiny energetic changes in the electron gas 
\cite{MarcusSpin} and for studying electron dephasing and inelastic
relaxation rates \cite{MarcusDephasing,Sivan}. A convenient object \cite
{MarcusOld}, once used as a mesoscopic thermometer \cite{MarcusT}, consists
of a lateral semiconductor dot weakly coupled to the reservoirs via two
leads, $l$ and $r$, each with $N_{l,r}\gtrsim 1$ open conducting channels,
and, therefore, quantum conductances $g_{l,r}=\frac{2e^{2}}{h}N_{l,r}$.
Information concerning fine energetic characteristics of single-particle
electron states in a dot can be extracted from the variance and parametric
correlations of universal conductance fluctuations (UCF), $\delta
g=g-\langle \langle g\rangle \rangle $, measured as random oscillations of
the dot conductance, $g$ around the mean value, $\langle \langle g\rangle
\rangle =g_{l}g_{r}/(g_{l}+g_{r})$, upon variation of a perpendicular
magnetic field \cite{Washburn,AltshulerKhm}, the Fermi energy \cite
{ChemicalPot,AltshulerKhm} or the dot shape \cite{MarcusOld}.

Energetic resolution of such a spectroscopy is set by the level broadening
of single-particle states in a particular device, which is limited by the
carrier escape into the leads, 
\[
\tau _{{\rm esc}}^{-1}=(N_{l}+N_{r})\Delta /h,
\]
where $\Delta =2\pi \hbar ^{2}/mS$ \ is the mean level spacing of
single-particle states of spin-polarized electrons with mass $m$ in a dot
with area $S$. The use of larger dots with weaker coupling to the leads
increases the sensitivity of the dot conductance to the variation of
external parameters. The use of larger dots also enables one to assess
directly the low excitation energy characteristics of the 2D electron gas,
since the electron properties in 1$\div $10$\mu $m$^{2}$-area dots
containing $10^{3}-10^{4}$ particles are less affected by the confinement
effects. Recently, large area dots were used for studying spin-polarisation
of a 2D electron gas \cite{MarcusSpin}. In order to enhance coupling between
a magnetic field and electron spin, J.Folk {\it et al} \cite{MarcusSpin}
used a magnetic field finely tuned to be orientated exactly parallel to the
plane of 2D electrons. One observed a strong suppression of the variance $%
\langle \langle \delta g^{2}(B_{\Vert })\rangle \rangle $ by an in-plane
field $B_{\Vert }$ interpreted in terms of a spin-orbit coupling in the 2D
gas \cite{Halperin}. This feature has also been accompanied by an
observation \cite{MarcusUnpublished} of such a negative weak localisation
(WL) magnetoresistance caused by an in-plane field $B_{\Vert }$ that one
would relate to the time-reversal symmetry breaking in the orbital motion of
electrons. In the present publication, we assess to which extent one can
reduce the influence of an ideally in-plane tuned magnetic field on quantum
transport in lateral semiconductor dots to spin effects alone, that is, we
determine the range of fields $B_{\Vert }$ that would affect WL and UCF's in
experimentally studied devices \cite{MarcusSpin}, in addition to
spin-related phenomena.

The influence of an in-plane magnetic field on the orbital motion of
carriers in a heterostructure or quantum well is the result of a finite
width, $\lambda _{z}$ of a 2D layer, and it has been discussed previously in
various contexts \cite{AnisMass,Spectroscopy}. The Lorentz force generated
by a planar field on electrons moving across $\vec{B}_{\Vert }$ within the
2D plane mixes up the electron motion along and across the confinement
direction, thus resulting in the electron momentum, $\vec{p}$ dependent
subband mixing and, therefore, in a modification of the 2D dispersion, $%
E(p)\rightarrow E(\vec{B}_{\Vert },\vec{p})$. In particular, (a) the 2D
electron mass increases in the direction perpendicular to $\vec{B}_{\Vert }$%
, whereas (b) in heterostructures which have no inversion symmetry in the
form of confining potential, $\vec{B}_{\Vert }$ also lifts the $\vec{p}%
\rightarrow -\vec{p}$ symmetry in the dispersion law \cite{AnisMass}: $E(%
\vec{B}_{\Vert },\vec{p})-E(\vec{B}_{\Vert },-\vec{p})\propto $ ($\vec{p}%
\cdot \lbrack \vec{B}_{\Vert }\times \vec{l}_{z}])^{3}\neq 0$.

The latter change in dispersion has potential to reduce the fundamental
symmetry of chaotic dot from orthogonal ($o$) to unitary ($u$), as a
perpendicular magnetic field would do.\ Below, we show that the efficiency
of time-reversal symmetry breaking by an in-plane magnetic field can be
characterized using the rate $\tau _{B\Vert }^{-1}\sim bB_{\Vert
}^{6}+aB_{\Vert }^{2}$ described in Eq. (\ref{TauB}). A similar conclusion
has recently been made in Ref. \cite{Meyer}. Without spin-orbit effects,
this parameter would determine the value of the WL correction, $g_{{\rm WL}%
}(B_{\Vert })\equiv \langle \langle g(B_{\Vert })\rangle \rangle -\langle
\langle g\rangle \rangle _{u}$ and of the variance of UCF, $\langle \langle
\delta g^{2}(B_{\Vert })\rangle \rangle $, as compared to their nominal
values, $g_{{\rm WL}}(0)\equiv \langle \langle g\rangle \rangle _{o}-\langle
\langle g\rangle \rangle _{u}$ and $\langle \langle \delta g^{2}\rangle
\rangle _{u}$ \cite{Pluhar,Efetov}: 
\begin{eqnarray}
g_{{\rm WL}}(B_{\Vert }) &=&g_{{\rm WL}}(0)\left[ 1+\tau _{B\Vert
}^{-1}/\tau _{{\rm esc}}^{-1}\right] ^{-1};  \label{Eq1} \\
\langle \langle \delta g^{2}(B_{\Vert })\rangle \rangle  &=&\langle \langle
\delta g^{2}\rangle \rangle _{u}\left\{ 1+\left[ 1+\tau _{B\Vert }^{-1}/\tau
_{{\rm esc}}^{-1}\right] ^{-2}\right\} .  \nonumber
\end{eqnarray}
The latter parameters can be studied from the UCF's fingerprints measured in
the 'shape of a dot' space in multi-gate devices \cite{MarcusOld}, or by
varying the Fermi energy in back-gated dots. The rise in the mass anisotropy
upon the increase of $B_{\Vert }$ would also manifest itself: as a change in
the UCF pattern. A varying dispersion relation for electrons studied at
different fields, $B_{\Vert 1}$ and $B_{\Vert 2}$, can be characterized
using the rate $\tau _{{\rm d}}^{-1}(B_{\Vert 1},B_{\Vert 2})\propto
(B_{\Vert 1}^{2}-B_{\Vert 2}^{2})^{2}$ in Eq.(\ref{TauD}), which can be used
to describe auto-correlation properties of a full $B_{\bot }$-dependent UCF
pattern, 
\begin{equation}
\frac{\langle \langle \delta g(B_{\Vert 1})\delta g(B_{\Vert 2})\rangle
\rangle _{u}}{\langle \langle \delta g^{2}\rangle \rangle _{u}}=\left[ 1+%
\frac{\tau _{{\rm d}}^{-1}(B_{\Vert 1},B_{\Vert 2})}{\tau _{{\rm esc}}^{-1}}%
\right] ^{-2}.  \label{Eq2}
\end{equation}

In the presence of an in-plane magnetic field, the effective 2D Hamiltonian
for electrons in a heterostructure with a potential profile $V(z)$ can be
obtained from the 3D Hamiltonian,

\begin{equation}
{\rm \hat{H}}_{3D}=-\frac{\hbar ^{2}\partial _{z}^{2}}{2m}+V(z)+\frac{%
(-i\hbar \nabla -\frac{e}{c}\vec{A})^{2}}{2m}+u(\vec{r},z),  \label{H3d}
\end{equation}
using the plane wave representation, $\Psi _{\vec{p}}=e^{i\vec{p}\vec{r}%
/\hbar }\varphi _{0\vec{p}}(z)$ for the lowest subband electrons. Here, $%
\vec{A}=(z-z_{0})\vec{B}_{\Vert }\times \vec{l}_{z}$ is the vector
potential, $z_{0}=$ $\langle 0|z|0\rangle $ is the center of mass position
of the electron wave function $|0\rangle \equiv \varphi _{0}^{(0)}(z)$ in
the lowest subband for $B_{\Vert }=0$, and $u(\vec{r},z)$ is a combination
of Coulomb potential of impurities and lateral potential forming the quantum
dot. Due to mixing between subbands $|0\rangle $ and $|n>0\rangle $ by an
in-plane magnetic field, $z$-dependent components $\varphi _{0\vec{p}}(z)$
are different for different in-plane momenta, $\vec{p}$, and we use both the
perturbation theory analysis \cite{PTresult} and a numerical
self-consistent-field technique to find $\varphi _{0\vec{p}}(z)$ and the
energy $E(\vec{B}_{\Vert },\vec{p})$ for each plane wave state.

For a weak or intermediate-strength magnetic field $\vec{B}_{\Vert }$, the
effective 2D Hamiltonian takes the form 
\begin{equation}
{\rm \hat{H}}_{2D}=\frac{\vec{p}^{2}}{2m}-p_{\bot }^{2}\gamma (B_{\Vert
})+p_{\bot }^{3}\beta (B_{\Vert })+u(\vec{r}).  \label{H2D}
\end{equation}
In Eq.(\ref{H2D}), $\vec{p}=-i\hbar \nabla $ $-\frac{e}{c}\vec{a}(\vec{r})$
is a purely 2D momentum operator, and $p_{\bot }=\vec{p}\cdot \lbrack \vec{B}%
_{\Vert }\times \vec{l}_{z}]/B_{\Vert }$ is its component perpendicular to $%
\vec{B}_{\Vert }$. Two additional terms in the free electron dispersion part
of ${\rm \hat{H}}_{2D}$ are the result of the $p_{\bot }$-dependent
inter-subband mixing. The first of them lifts rotational symmetry by causing
an anisotropic mass enhancement \cite{AnisMass}. It increases the 2D density
of states and, for a 2D gas with a fixed sheet density, it reduces the Fermi
energy calculated from the bottom of the 2D conduction band, $E_{{\rm F}%
}(B_{\Vert })=E_{{\rm F}}^{0}-\frac{\gamma (B_{\Vert })}{2}p_{{\rm F}}^{2}$.
A cubic term in ${\rm \hat{H}}_{2D}$ is related \ to the time-reversal
symmetry breaking by $B_{\Vert }$. Note that, depending on the choice of a
gauge, one may also generate a linear $p_{\bot }$ term, but this one can be
eliminated by a trivial gauge transformation. A perturbation theory analysis
of this problem is discussed in footnote \cite{PTresult}, and for a moderate
field it results in the parametric dependences $\gamma \sim m^{-1}\left(
\lambda _{z}/\lambda _{B}\right) ^{4}$ and $\beta \sim (\lambda _{z}/m\hbar
)\left( \lambda _{z}/\lambda _{B}\right) ^{6}$.

\begin{figure}
\hspace{0.05\hsize}
\epsfxsize=0.78\hsize
\epsffile{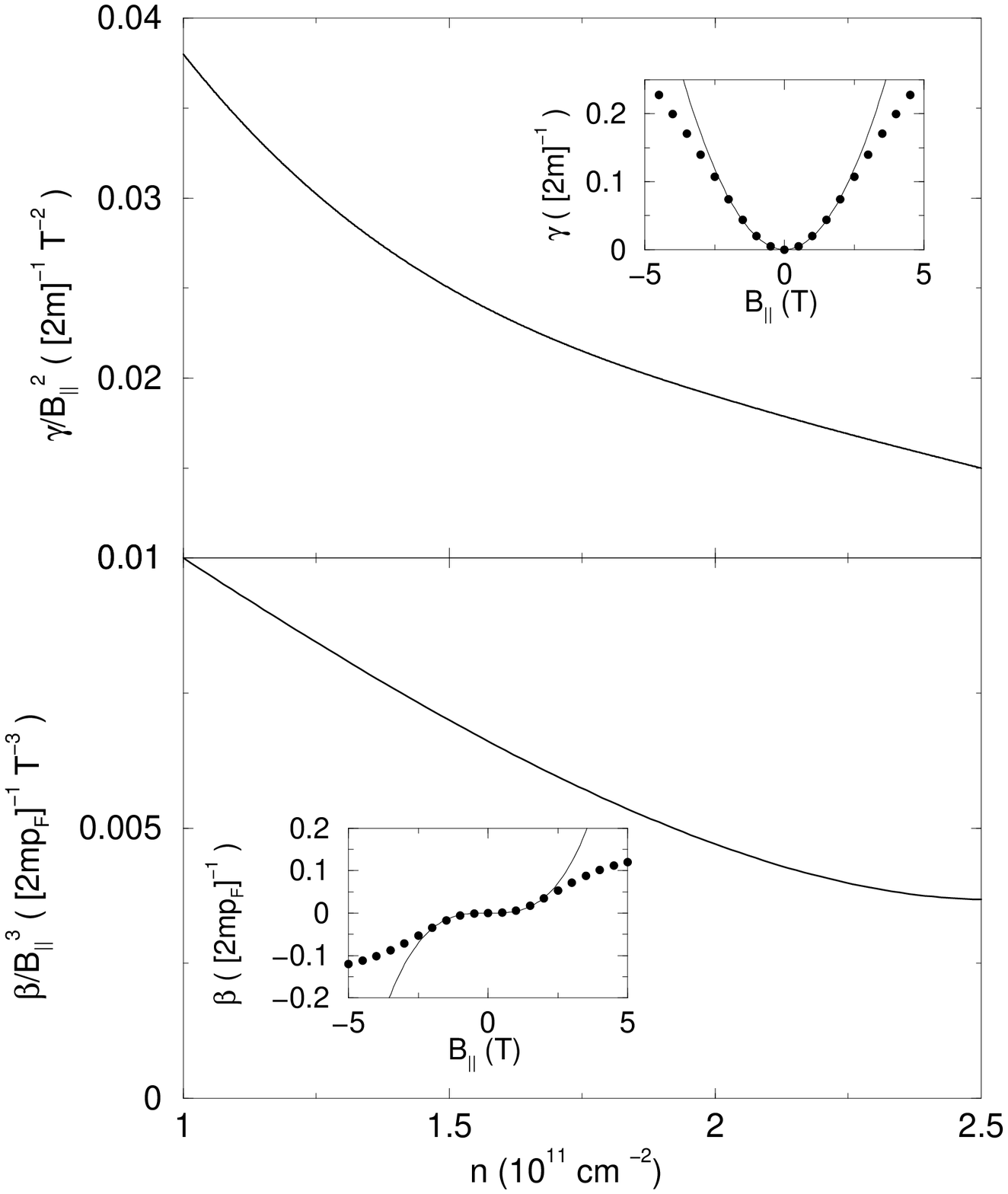}
\refstepcounter{figure}
\label{FigFJ1}

{\setlength{\baselineskip}{10pt} FIG.\ 1.
Calculated dependence of parameters $\protect\gamma$ and $\protect%
\beta$ on the sheet density of 2D electrons. Insets show the effect of $%
B_{||}$ on symmetric, $[E(p_{\bot })+E(-p_{\bot })]/2$ and anti-symmetric, $%
[E(p_{\bot })-E(-p_{\bot })]/2$ parts of the 2D electron dispersion in a
braoder range of $B_{||}$, where a pertubative expansion is not applicable. }
\end{figure}

To obtain quantitative estimates for parameters $\gamma $ and $\beta $, we
evaluated the electron dispersion at in-plane magnetic fields using a full
self-consistent numerical method. The quantum well confining potential $V(z)$
was constructed using the nominal growth parameters of the sample studied in 
\cite{MarcusSpin} which was a Al$_{.34}$Ga$_{.66}$As/GaAs heterojunction. $%
V(z)$ also included Hartree and exchange-correlation potentials generated by
the free carriers in the quantum well. The Hartree potential was derived
from the $z$-dependent 3D density of electrons by numerical solution of the
Poisson equation. The exchange-correlation term was calculated within the
local-density approximation \cite{vosko}. A flat-band boundary condition was
used, i.e., we assumed that the electric field produced by donors in the
(Al,Ga)As barrier is screened out in the GaAs buffer-layer by the 2D
electron gas. In each loop of the self-consistent procedure we solved
numerically the Schr\"{o}dinger equation with the Hamiltonian in Eq.(\ref
{H3d}) to get the 3D electron density, neglecting $u(\vec{r},z)$. Then, a
new $V(z)$ was constructed, which entered the next loop of the procedure
until the self-consistency condition was achieved. The numerically obtained
dependences of $\gamma $ and $\beta $ on the in-plane magnetic field for an
electron sheet density of 2$\times $10$^{11}$~cm$^{-2}$ are shown in insets
of Fig.\ref{FigFJ1} a) and b). At low fields, $\gamma \sim B_{\parallel }^{2}
$ and $\beta \sim B_{\parallel }^{3}$, as anticipated in the perturbation
theory treatment. The proportionality coefficients are plotted in Fig.\ref
{FigFJ1} a) and b) versus the electron sheet density. Both the effective
mass renormalization in the quadratic term of energy dispersions and the
time-reversal symmetry breaking cubic term are larger at lower 2D electron
gas densities, due to a weaker confining electric field ({\it i.e.}, longer $%
\lambda _{z}$).

In ${\rm \hat{H}}_{2D}$ in Eq. (\ref{H2D}), disorder is incorporated in the
form of a scattering potential $u(\vec{r})\approx $ $\langle 0|u(\vec{r}%
,z)|0\rangle $. This can be characterized by the value of the mean free
path, $l\gg h/p_{{\rm F}}$, or a momentum relaxation time $\tau $ related to
the diffusion coefficient $D=v_{{\rm F}}^{2}\tau /2$. The modification of
the electron density of states by $B_{\Vert }$ only slightly affects the
value of the electron mean free path. The presence of a parallel field also
changes the symmetry of Born amplitudes of scattering between plane waves $%
\Psi _{\vec{p}}=e^{i\vec{p}\vec{r}/\hbar }\varphi _{0\vec{p}}(z)$, $f_{\vec{p%
}\vec{p}^{\prime }}=\langle \Psi _{\vec{p}}|u(\vec{r},z)|\Psi _{\vec{p}%
^{\prime }}\rangle $. Due to the momentum-dependent subband mixing, $f_{\vec{%
p}\vec{p}^{\prime }}$ acquires an addition, $f_{\vec{p}\vec{p}^{\prime }}=f_{%
\vec{p}\vec{p}^{\prime }}^{(0)}\left\{ 1+(p_{\bot }+p_{\bot }^{\prime
})B_{\Vert }\zeta \right\} $, where $\zeta =\frac{e}{mc}\sum_{n\geq 1}\frac{%
\langle 0|u(\vec{p}-\vec{p}^{\prime },z)|n\rangle }{\langle 0|u(\vec{p}-\vec{%
p}^{\prime },z)|0\rangle }\frac{\langle n|z|0\rangle }{\varepsilon
_{n}-\varepsilon _{0}}$, which is equivalent to the presence of a random
gauge field in the effective 2D Hamiltonian \cite{Falko90,Mathur}, 
\[
\vec{a}=2[\vec{B}_{\Vert }\times \vec{l}_{z}]\sum_{n\geq 1}\frac{\langle 0|u(%
\vec{r},z)|n\rangle \langle n|z|0\rangle }{\varepsilon _{n}-\varepsilon _{0}}%
. 
\]
The latter can be interpreted as a result of an effective 'curving' of a 2D
plane by impurities in systems with $z$-dependent scattering potential,
which in the presence of an in-plane magnetic field\ generates a random
effective perpendicular field component, $b_{\bot }=\left[ {\rm rot}\vec{a}%
\right] _{z}$. In systems, where scattering is dominated by Coulomb centers
behind a spacer and is almost independent of $z$, a smaller effect may be
taken into account, $\delta \vec{a}=\eta \lbrack \vec{B}_{\Vert }\times \vec{%
l}_{z}]([\vec{B}_{\Vert }\times \vec{l}_{z}]\cdot \nabla )^{2}u(\vec{r})$.
However, $\delta \vec{a}$ has a negligible influence on the quantum
transport characteristics of 2D electrons, as compared to the effect of
dispersion.

In the present paper, we study quantum transport characteristics of chaotic
dots by modelling them as 2D disordered billiards with a short-range
disorder. It has been shown before that the results obtained for a
zero-dimensional limit of diffusive systems, $\tau _{{\rm esc}}\gg L^{2}/D$
and $L>l$, are universally applicable to the description of WL and UCF's in
a broad variety of quantum chaotic billiards\ \cite{EfetovBook}, even in
ballistic ones \cite{Baranger}. \ We also used a semiclassical diagrammatic
language to calculate two-particle correlation functions, Cooperons $P_{{\rm %
C}}(\omega ;\vec{R},\vec{R}^{\prime })$ and diffusons $P_{{\rm d}}(\omega ;%
\vec{R},\vec{R}^{\prime })$ \cite{AltshulerKhm}. These correlation functions
emerge in the form of ladder diagrams from the perturbation theory analysis
upon averaging over disorder the Kubo-formula conductance. Schematically,
the form of a weak localisation correction and of the variance and
correlation function of UCF can be represented as $g_{{\rm WL}}(B_{\Vert
})\propto \int d\vec{R}W(\vec{R})P_{{\rm C}}(0;\vec{R},\vec{R})$ and $%
\langle \langle \delta g(B_{\Vert 1})\delta g(B_{\Vert 2})\rangle \rangle
\propto $ $\ \int d\vec{R}d\vec{R}^{\prime }W(\vec{R})W(\vec{R}^{\prime
})\sum_{{\rm d},{\rm C}}|P_{{\rm d},{\rm C}}(\omega ;\vec{R},\vec{R}^{\prime
})|^{2}$, where $\omega =E_{{\rm F}}(B_{\Vert 1})-E_{{\rm F}}(B_{\Vert 2})$,
and $E_{{\rm F}}(B_{\Vert })$ is the Fermi energy of the 2D gas calculated
from the bottom of the 2D conduction band determined in Eq.(\ref{H2D}).
Dispersionless weight factors $W(\vec{R})$ both take care of the particle
number conservation upon diffusion inside a dot \cite{UCF-ends} and
incorporate coupling parameters to the leads. In the zero-dimensional limit,
both $g_{{\rm WL}}$ and $\langle \langle \delta g(B_{\Vert 1})\delta
g(B_{\Vert 2})\rangle \rangle $ are dominated by the lowest Cooperon
(diffuson) relaxation mode, $\tau _{{\rm esc}}^{-1}+\tau _{{\rm C}({\rm d}%
)}^{-1}$ determined by the interplay of the escape to the reservoirs and the
Cooperon (diffuson ) suppression by time-reversal symmetry breaking (the
difference in condition of quantum diffusion). The latter effect can be
analyzed for an infinite system, where the derivation is simplified by the
use of Fourier representation for Cooperons (diffusons).

Using the Fourier form,\ the equation for the diffuson (Cooperon)
correlation function,\ $\Pi \cdot P_{{\rm d}({\rm C})}(\omega ,\vec{q})=\tau
^{-1}$, can be obtained from the analysis of a kernel ($\hbar =1$), 
\[
\Pi (\omega ,\vec{q})=1-\int \frac{d\vec{p}}{2\pi \nu \tau }G_{B_{\Vert 1}}^{%
{\rm R}}(\varepsilon ,\vec{p})G_{B_{\Vert 2}}^{{\rm A}}(\varepsilon -\omega
,\pm \lbrack \vec{p}-\vec{q}]),
\]
where sign '+/-' is related to diffuson (Cooperon), respectively.
Disorder-averaged retarded and advanced single-particle Green functions, $G^{%
{\rm R,A}}$ correspond to different values of $B_{\Vert }$. $G^{R,A}$ were
calculated pertubatively with respect to all terms containing $B_{\Vert }$,
which relies on the assumption that within the relevant parametric regime
the variation of the energy, $\delta E(p_{{\rm F}})$, induced by $B_{\Vert }$
is small in comparison with the scattering rate, $\delta E\ll h/\tau $. The
result has the form of the diffusion equation,

\begin{equation}
\left[ -i\{\omega +\delta \}-D\nabla ^{2}+\tau _{{\rm d}}^{-1}\right] P_{%
{\rm d}}=\delta (\vec{R}-\vec{R}^{\prime }).  \label{diffusonEq}
\end{equation}
It contains $\tilde{\omega}=\omega +\delta $ with $\delta =p_{{\rm F}%
}^{2}[\gamma (B_{\Vert 1})-\gamma (B_{\Vert 2})]/2\hbar $ and the rate 
\begin{equation}
\tau _{{\rm d}}^{-1}=\frac{\tau p_{{\rm F}}^{4}}{8\hbar ^{2}}[\gamma
(B_{\Vert 1})-\gamma (B_{\Vert 2})]^{2}+\frac{2\zeta ^{2}}{\tau }p_{{\rm F}%
}^{2}[B_{\Vert 1}-B_{\Vert 2}]^{2}  \label{TauD}
\end{equation}
The first term in Eq.(\ref{TauD}) comes from the deformation of a Fermi
circle by $B_{\Vert }$, the second takes into account the field effect upon
the scattering of plane waves. Eq.(\ref{diffusonEq}) also contains the
difference between the electron kinetic energies in two measurements of
conductance, $\omega =E_{{\rm F}}(B_{\Vert 1})-E_{{\rm F}}(B_{\Vert 2})$,
each of them shifted, $E_{{\rm F}}(B_{\Vert })=E_{{\rm F}}^{0}-\frac{1}{2}p_{%
{\rm F}}^{2}\gamma (B_{\Vert })$ with respect to the Fermi energy, $E_{{\rm F%
}}^{0}$ in the electron gas with the same sheet density at $B_{\Vert }=0$.\
The latter fact is important, since, for lateral dots where electron density
is fixed, one should substitute $\tilde{\omega}=\omega +\delta =0$, so that
only the $B_{\Vert }$-dependent anisotropy of the electron wavelength along
the Fermi line affects the interference pattern of current carriers. \ 

The Cooperon equation derived after the calculation of the integral in $\Pi
(\omega ,\vec{q})$ can be represented in the form

\begin{equation}
\left[ -i\tilde{\omega}+D(-i\nabla -\vec{q})^{2}+\tau _{{\rm C}}^{-1}+\tau
_{d}^{-1}\right] P_{C}=\delta (\vec{R}-\vec{R}^{\prime }).
\label{CooperonEq}
\end{equation}
It contains an additional decay rate, $\tau _{{\rm C}}^{-1}(B_{\Vert
1},B_{\Vert 2})$, 
\begin{eqnarray}
\tau _{{\rm C}}^{-1} &=&\frac{\tau p_{{\rm F}}^{6}}{8\hbar ^{2}}\left[ \frac{%
\beta (B_{\Vert 1})+\beta (B_{\Vert 2})}{2}\right] ^{2}+\frac{\zeta ^{2}p_{%
{\rm F}}^{2}}{2\tau }[B_{\Vert 1}+B_{\Vert 2}]^{2}  \nonumber \\
\tau _{B\Vert }^{-1} &=&\tau _{{\rm C}}^{-1}(B_{\Vert },B_{\Vert }),
\label{TauB}
\end{eqnarray}
which accounts for dephasing between electrons encircling the same chaotic
trajectory in reverse directions, the result of lifting the time-reversal
symmetry by an in-plane magnetic field. This parameter is relevant for
describing WL correction in Eq. (\ref{Eq1}), and also the $o\rightarrow u$
crossover in conductance fluctuations.

To mention, a shift in the Cooperon gauge in Eq.(\ref{CooperonEq}), $\vec{q}=%
\frac{3}{2\hbar }p_{{\rm F}}^{2}m\beta (B_{\Vert })[\vec{B}_{\Vert }\times 
\vec{l}_{z}]/B_{\Vert }$, is the result of the following artifact: cubic
term in the effective electron dispersion not only lifts the inversion
symmetry of the line $E(B_{\Vert },\vec{p})=E_{{\rm F}}(B_{\Vert })$, but
also shifts its geometrical centre with respect to the true bottom of the 2D
conduction band. Since in conductance calculations only electrons with $E=E_{%
{\rm F}}$ matter, such a shift would be eliminated by choosing a slightly
modified initial gauge, which can now be corrected by applying a gauge
transformation $P_{{\rm C}}=e^{i\vec{q}\vec{R}}\tilde{P}_{{\rm C}}$ directly
to the Cooperon. One may say that, here, the phase-coherent transport is
only affected by the $B_{\Vert }$-induced ($\vec{p}\rightarrow -\vec{p}$%
)-asymmetric distortion of the Fermi circle into an oval, but not by a shift
of such an oval in the momentum space.

The effect of time-reversal symmetry breaking by the in-plane field
described in Eq.(\ref{TauB}) has a field dependence $\tau _{B\Vert
}^{-1}\sim bB_{\Vert }^{6}+aB_{\Vert }^{2}$. In a ballistic billiard, or in
a heterostructure with $z$-independent scattering potential, it would be
dominated by the unusual $B_{\Vert }^{6}$-dependence (also found in ref.  
\cite{Meyer}) that we attribute to the effect of cubic term generated by $%
B_{\Vert }$ in the 2D electron dispersion, thus giving rise to a relatively
sharp crossover between 'flat' regions related to orthogonal and unitary
symmetry regimes. For a large-area (8$\mu $m$^{2}$) quantum dot with
electron density $2\times 10^{11}$cm$^{-2}$ studied by J.Folk {\it et al} 
\cite{MarcusSpin,MarcusUnpublished}, we estimated the crossover field as $%
B_{\Vert }=0.6\div 0.8$T. When studying the crossover, one has to take into
account that the in-plane field also causes fluctuations in conductance,
without breaking time-reversal symmetry, as described by Eqs. (\ref{Eq2},\ref
{TauD}). For the same parameters of a structure, we estimated the field
where such a random dependence would appear as $B_{\Vert }\sim 0.3$T, and
the result in Eq. (\ref{TauD}) suggests that, for a perfectly in-plane field
orientation, variation of the UCF fingerprint is faster at higher fields.

We thank B.Altshuler, C.Marcus and J.Meyer for discussions and for
information concerning the unpublished works \cite{MarcusUnpublished,Meyer}.
This research was funded by EPSRC, NATO CLG and EU COST programmes.

\begin{references}
\bibitem{PTresult}  Using perturbation theory expansion for the plane wave
energy, up to the third order in $p_{\bot }$ (assuming that $p_{{\rm F}%
}\langle n|z|0\rangle <p_{{\rm F}}\lambda _{z}<\hbar $), we arrived at $%
\delta E=\delta _{2}+\delta _{3}$, 
\begin{eqnarray*}
\delta _{2} &\approx &-\frac{p_{\bot }^{2}B_{\Vert }^{2}}{m^{2}}\sum_{n\geq
1}\frac{\langle n|z|0\rangle ^{2}}{\varepsilon _{n}-\varepsilon _{0}}\sim -%
\frac{p_{\bot }^{2}}{m}\left( \frac{\lambda _{z}}{\lambda _{B}}\right) ^{4}
\\
\delta _{3} &\approx &\frac{p_{\bot }^{3}B_{\Vert }^{3}}{m^{3}}\sum_{k,n\geq
1}\frac{\langle 0|z|k\rangle \langle k|z|n\rangle \langle n|z|0\rangle }{%
[\varepsilon _{k}-\varepsilon _{0}][\varepsilon _{n}-\varepsilon _{0}]}\sim 
\frac{p_{\bot }^{3}\lambda _{z}}{m\hbar }\left( \frac{\lambda _{z}}{\lambda
_{B}}\right) ^{6}
\end{eqnarray*}
where $\lambda _{B}=\sqrt{\hbar c/eB_{\Vert }}$ is the magnetic length.
\end{references}
\end{document}